SOLID-STATE PHYSICS

# Threes company

Benjamin J. Wieder

*Enabled by recent advances in symmetry and electronic structure, researchers have observed signatures of unconventional threefold degeneracies in tungsten carbide, challenging a longstanding paradigm in nodal semimetals.*

Take a single layer of graphite and you get graphene, a material whose structural and electronic properties allow diverse applications ranging from biosensing to electrical engineering. Try and explain graphene's properties using solid state physics, and you get an equation similar to one otherwise seen in discussions of cosmology and colliders: the Dirac Equation. In the decade following graphene's discovery, all materials of this kind, known as 'nodal semimetals', were named and categorized assuming a one-to-one correspondence between the low-energy electronic behavior of crystalline solids and the equations of high-energy particle physics, leading to the classification of a wide variety of chemical compounds as Dirac or Weyl semimetals. Now, writing in *Nature Physics*, Jun-Zhang Ma and colleagues [1] present experimental evidence in tungsten carbide of a semimetal that escapes this paradigm.

This assumption of one-to-one correspondence was first upended in early 2016 by a pair of papers [2, 3] noting that, in fact, plenty of nodal semimetals are instead governed by equations decidedly unlike those in high energy physics. These 'unconventional' semimetals, as characterized in a flood of subsequent theoretical proposals, feature electronic states twisted into loops, chains, and hourglasses, or meeting in unexpected multiples of three. Materials candidates accompanying these proposals were identified so readily that one might even question how exotic unconventional semimetals really are. Rather, it was possible that we had all just been a bit spoiled by the natural simplicity of graphene. By observing electronically relevant threefold nodal degeneracies and surface Fermi arcs in tungsten carbide [1], Ma et al. provide some of the first experimental support for the growing body of imaginative proposals for unconventional semimetals with topological electronic character.

In both particle physics and nodal semimetals, the possible energies of a particle or quasiparticle are determined by its momentum. This is known as a dispersion relation. When a particle has no mass, the solutions of its dispersion relation come together and meet in a degenerate nodal point. In high-energy physics, the structure of this dispersion relation and the degree of its degeneracy are determined by the fundamental symmetries of nature, such as charge conjugation, parity inversion, and time reversal. Conversely, as recognized since the 1970's [4], the dispersion relation of a crystalline solid is instead more accurately described by the mathematical representations of its spatial symmetries. With the advent of computers powerful enough to perform large-scale calculations of the electronic structures of real materials, researchers have only recently begun to link this abstract mathematical characterization to nodal points in known chemical compounds. In turn, this fueled a rediscovery of the group theory underlying the connection between symmetry and dispersion, culminating this past year in a complete symmetry-based characterization of all possible electronic structures in nonmagnetic crystals [5].

Among the proposals for unconventional nodal quasiparticles, the notion of a threefold electronic degeneracy was perhaps the most unexpected. Electrons are fermions, or particles with half-integer spin. Fermions are known to come in degenerate pairs under time-reversal symmetry, and so it was assumed that the nodal degeneracies of massless condensed-matter quasiparticles should manifest in multiples of two: either as twofold degenerate Weyl fermions, with broken time-reversal or parity symmetry, or fourfold degenerate Dirac fermions, with time-reversal and parity symmetry (Fig. 1a). Under this paradigm, states with broken symmetries could be fine-tuned to meet in multiples of three, but would have no reason to do so in real materials (Fig. 1b). However, utilizing the recent understanding of the constraints imposed by crystalline symmetry, two independent sets of viable proposals emerged for realizing a threefold electronic quasiparticle. One could search for unique symmetry combinations in cubic crystals that protect a threefold degenerate state with linear dispersion in all directions, a paradigm-challenging 'spin-1 Weyl fermion' (Fig. 1c) that comes accompanied by topologically protected surface states [2]. Or one could look for materials in which twofold degenerate states crossed with singly degenerate states (Fig. 1d), forming an intermediate quasiparticle between Dirac and Weyl [6-8].

Ma et al. [1] provide experimental evidence of this second kind of threefold fermion at energies relevant to transport in tungsten carbide. They also observe relatively robust Dirac-type Fermi arc surface states in this material. However, while these surface arcs are topological in a sense, they are more of a remnant of band inversion in this material, and are not unique to, or necessarily linked to, the particular threefold fermions observed in tungsten carbide [9]. An experimental realization of the topological Fermi arcs necessitated by condensed matter spin-1 Weyl fermions still remains elusive.

However, the future is very bright for materials discovery and engineering towards this end. Exploiting these advances in symmetry and electronic structure, researchers have just recently proposed the presence of closely related topological surface states and doubled threefold fermions in readily synthesizable crystals in the RhSi family [10,11]. Given the immense number of known crystalline compounds still uninvestigated for nodal fermions, other ideal materials candidates will surely follow. With the availability of new search algorithms guided by crystal symmetry and fueled by 21$^{st}$ century computing power, solid state physics may soon reach an era in which the unconventional has become commonplace.

Benjamin J. Wieder
Princeton University, Princeton, NJ, USA.
e-mail: bwieder@princeton.edu

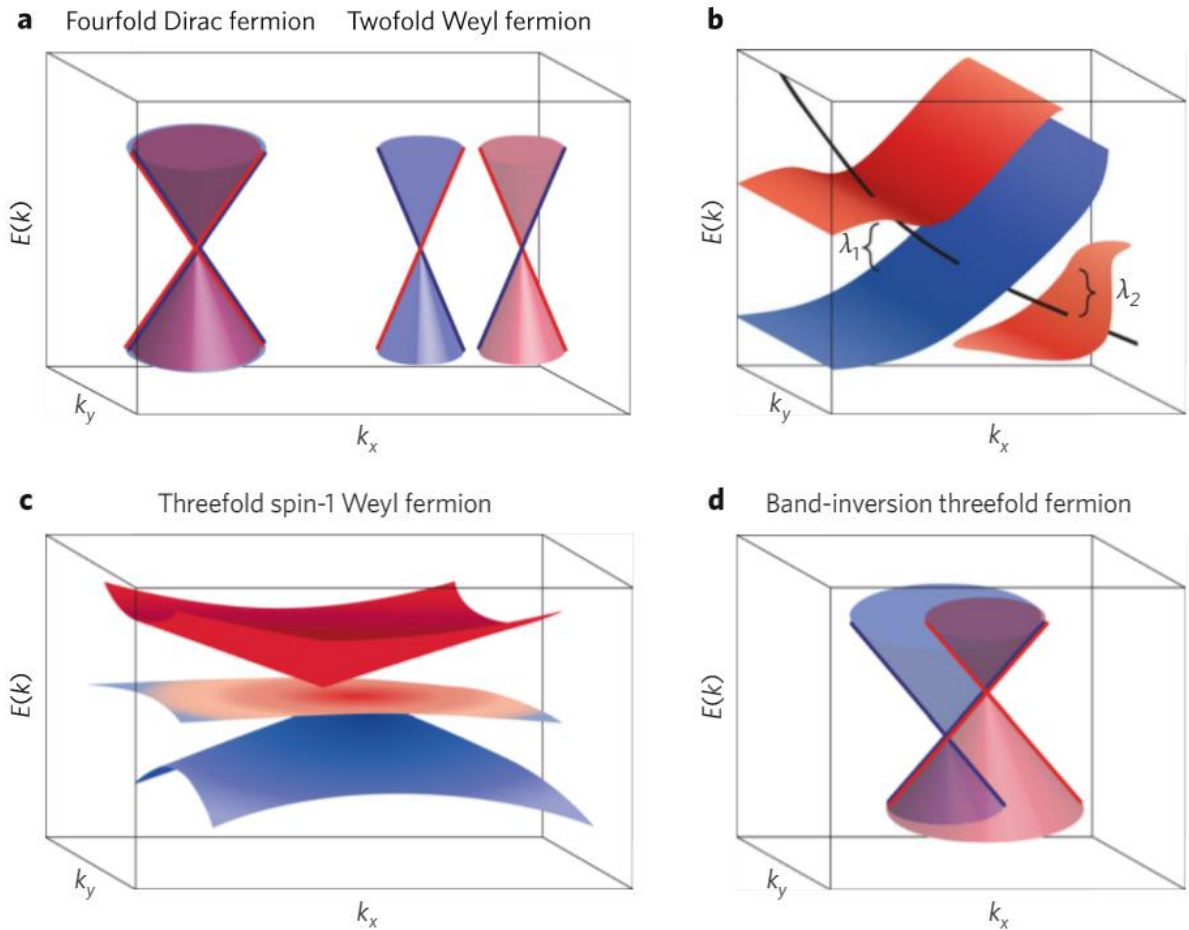

**Figure 1. Dispersion relations for electronic bands as a function of crystal momentum (k).** (a) The dispersion relation for a 3D Dirac nodal degeneracy. A Weyl fermion has the same dispersion, but with half the degeneracy. (b) For three arbitrary singly-degenerate bands, two independent parameters ($\lambda_{1,2}$) must be fine-tuned to get a threefold degeneracy. (c) A three-component spin-1 Weyl fermion has linear dispersion and single degeneracy in all three directions [2]. (d) A three-component band-inversion fermion, conversely, has a twofold degenerate line in one direction, and can be considered an intermediary between Dirac and Weyl fermions [1, 6-8].